\documentstyle[11pt,aaspp4,psfig]{article}

\received{2001}

\slugcomment{Submitted to The Astronomical Journal}

\newcommand{\src}{GCRT~J1746$-$2757}
\newcommand{\xtesrc}{XTE~J1748$-$288}
\newcommand{\mjybm}{\mbox{mJy~beam${}^{-1}$}}

\begin{document}

\title{Low-Frequency Radio Transients in the Galactic Center}

\author{Scott D.~Hyman}
\affil{Department of Physics, Sweet Briar College, Sweet Briar, VA
	24595; hyman@sbc.edu}

\author{T.~Joseph~W.~Lazio, Namir E.~Kassim}
\affil{Naval Research Laboratory, Code 7213,
	Washington, DC  20375-5320; lazio@rsd.nrl.navy.mil; kassim@rsd.nrl.navy.mil}

\begin{abstract}

We report the detection of a new radio transient source, \src, located
only 1\fdg1 north of the Galactic center.  Consistent with other radio
transients toward the Galactic center, this source brightened and
faded on a time scale of a few months.  No X-ray counterpart was
detected.  We also report new 0.33~GHz measurements of the radio
counterpart to the X-ray transient source, \xtesrc, previously
detected and monitored at higher radio frequencies.  We show that the
spectrum of \xtesrc\ steepened considerably during a period of a few
months after its peak.  We also discuss the need for a more efficient means of
finding additional radio transients.

\end{abstract}


\section{Introduction}

Known classes of highly variable and transient radio sources include
radio counterparts of X-ray sources and microquasars.  Although
there are many examples of variable radio sources discovered as a
result of high-energy observations, there are surprisingly few radio
surveys for highly variable or transient sources.  A radio
survey of the Galactic plane by Gregory \& Taylor~(1981, 1986) discovered 4 variable sources including GT~0236$+$610, a
Galactic X-ray binary, and 1 candidate transient.  The MIT-Green Bank
surveys (\cite{lhclcb90}; \cite{glhclb90};
\cite{ghcbl91}) discovered a number of variable sources ($<
40$\% variable).  An on-going program at NRAO Green Bank monitors the
Galactic plane at~8.4 and~14.4~GHz (\cite{lmdekz00}).

The Galactic center (GC) is a promising region in which to search for
highly variable and transient sources.  The stellar
densities are high, and neutron star- and black hole binaries appear
as (transient or variable) X-ray sources concentrated
toward the GC (\cite{gks93}). Previous surveys have been ill-suited
for detecting radio transients
toward the GC, however. Typically, they have utilized either single
dish instruments, which suffer from confusion in the inner Galaxy, or
they have utilized the VLA\footnote{The
Very Large Array (VLA) is a telescope of the National Radio Astronomy Observatory (NRAO) which is
operated by Associated Universities, Inc., under a cooperative agreement
with
the National Science Foundation.} for only a single epoch (e.g.,
\cite{zhbwp90}; \cite{bwhz94}).

The first two radio transients detected toward the GC were A1742$-$28
(\cite{dwben76}) and the Galactic Center Transient (GCT,
\cite{zhaoetal92}).  These two transients had similar radio
properties, but only the former was associated with an X-ray source.
More recently, radio counterparts to the X-ray transients,
\xtesrc\ (\cite{hrm98a}; \cite{hrgwm98b}; \cite{rhm98}) and
GRS~1739$-$278 (\cite{hrmmr96}) have been detected in the \hbox{GC}.

This paper reports the detection of a new transient radio source,
\src, toward the \hbox{GC}; it  also reports additional measurements of the 
radio counterpart to \xtesrc.  In \S\ref{sec:observe}, we
describe the observations that led to the discovery of \src, and in
\S\ref{sec:discuss} we describe and discuss what constraints we can
put on these two sources from our observations.  We conclude in
\S\ref{sec:conclude} with a brief discussion of future plans to monitor
the \hbox{GC} for transient and variable radio
sources.

\section{Observations}\label{sec:observe}

We have used recent advances in low-frequency and 3-dimensional
imaging to produce a wide-field image ($\approx 2\fdg5$) of the GC,
with uniform and high resolution across the field, from 0.33~GHz VLA
observations made between~1986 and~1989 (\cite{lklh00}).  A second
image, designed to exploit the increase in number and quality of the
0.33~GHz receivers on the VLA since the acquisition of the data used
by LaRosa et al.~(2000), is in progress and will be reported elsewhere
(Nord et al.~2002, in preparation).  However, these two epochs are
already sufficient to begin a search for variable and transient radio
sources.

Table~\ref{tab:observe} summarizes the relevant observations.  The
observations that form our new GC imaging program were obtained
in~1998.  The data were calibrated in the standard fashion (similar to
that of \cite{lklh00}), with 3C~48 and 3C~286 used to calibrate the
visibility amplitudes and frequent observations of the VLA calibrators
B1822$-$096 and B1830$-$360 used to calibrate the visibility phases.
The observations were made in spectral-line mode in order to permit
efficient excision of radio frequency interference and to reduce
bandwidth-smearing effects during imaging, with Cygnus~A used to
calibrate the bandpass. Wide-field, 3-dimensional imaging (via the
\textsc{aips} task \texttt{IMAGR}) was used to compensate for the
non-coplanar geometry of the \hbox{VLA} (\cite{cp92}).

\begin{deluxetable}{lccccc}
\tablecaption{Database of Radio Observations\label{tab:observe}}
\footnotesize
\tablehead{
 \colhead{Epoch} & \colhead{VLA Config.} & \colhead{Frequency} & \colhead{Bandwidth} & \colhead{Integration Time} & \colhead{RMS Noise} \\
                 &                       & \colhead{(GHz)} &
	\colhead{(MHz)} & \colhead{(hr)} & \colhead{(\mjybm)}
}
\startdata
1996~October~19 & A   & 0.33 & 2.9 & 5.8 & 35 \\
1996~October~24 & A   & 1.4  & 25  & 0.4 & 7  \\
1998~March~14   & A   & 0.33 & 3.1 & 5.6 & 20 \\
1998~September~25 & B & 0.33 & 3.1 & 6.7 & 11 \\
1998~November~29 & C  & 0.33 & 3.1 & 6.3 & 52 \\
2000~July~15,~19 & C/D & 1.42 & 50 & 1.4 & 6  \\
2000~July~15,~19 & C/D & 4.86 & 50 & 1.1 & 0.15 \\
2000~December~11 & A  & 1.42 & 50  & 2.3 & 0.08 \\
2000~December~11 & A  & 4.86 & 50  & 1.2 & 0.03 \\
\enddata
\tablecomments{We report only the new measurements.  See LaRosa et
al.~(2000) for the description of the 1986--1989 observations.}
\end{deluxetable}

As in the analysis of LaRosa et al.~(2000), small-diameter sources
were identified visually.  Approximiately 100 sources were detected
with peak intensity to background ratios greater than 5$\sigma$.

We identified one source on a 1998 image (\src, \S\ref{sec:trans})
that did not appear in the image of LaRosa et al.~(2000).  Additional
1.42 and~4.86~GHz continuum observations of this source were acquired
in~2000 July and December.  The data were calibrated and imaged in the
standard fashion, with 3C~48 used to calibrate the visibility
amplitudes and frequent observations of B1751$-$253 used to calibrate
the visibility phases.

In addition to the observations in~1998 and~2000, we shall also refer
to~0.33 and~1.4~GHz observations acquired in~1996 as part of a program
to search for pulsars in the \hbox{GC}.  Those observations shall be
reported fully elsewhere (Lazio \& Cordes~2002, in preparation), but
the calibration and imaging of them are similar to that described
above.

\section{Results and Discussion}\label{sec:discuss}

In this section we discuss two sources detected during our 1998
observations but that were not present in the LaRosa et al.~(2000)
image.  Anticipating that future work will find more radio transients
associated with the GC (\S\ref{sec:conclude}), we adopt a
nomenclature consistent with that recommended by the IAU (Lortet,
Borde, \& Ochsenbein~1994) for our previously undetected transient.
Namely, we describe it as a GCRT, Galactic Center Radio Transient.

\subsection{\src}\label{sec:trans}

\begin{figure}
\begin{center}
\mbox{\psfig{file=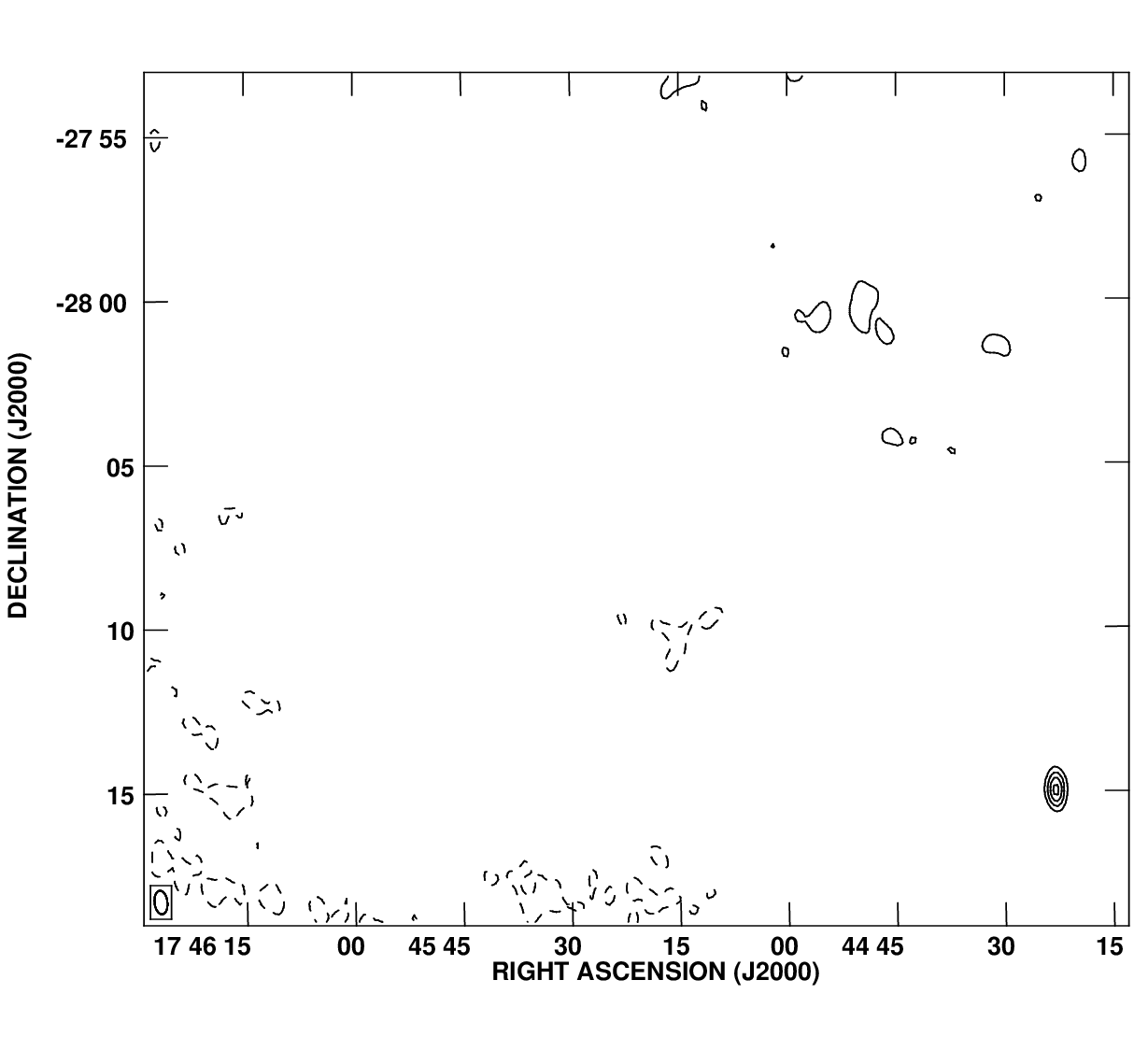,width=0.47\textwidth,angle=-90,silent=} \psfig{file=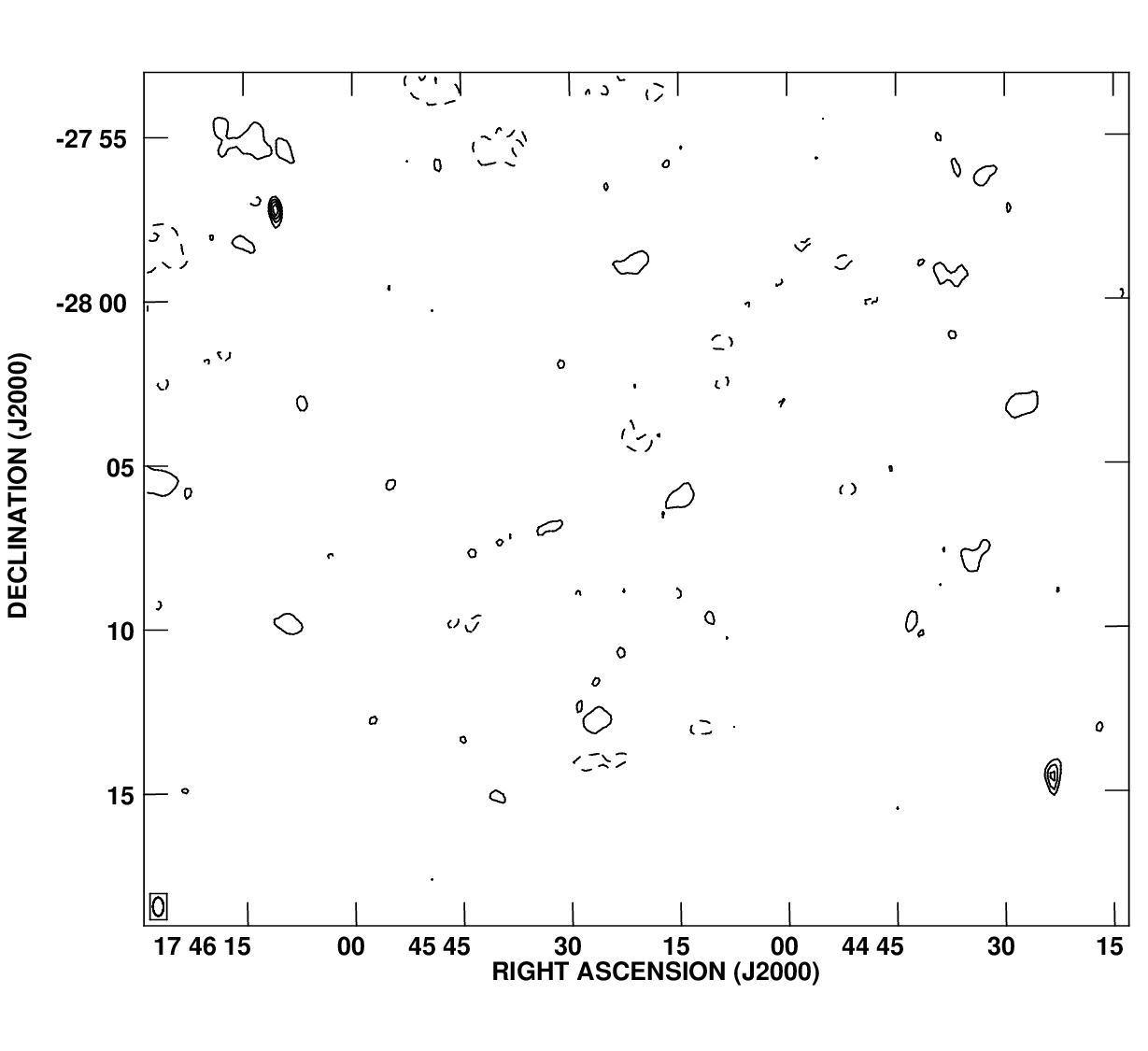,width=0.47\textwidth,angle=-90,silent=}}
\end{center}
\vspace*{-1cm}
\caption{The field containing the GC radio transient, \src, at~0.33~GHz.
(a)~Late 1980s from LaRosa et al.~(2000). The resolution is $43\arcsec \times
24\arcsec$.  The rms noise level is 10~\mjybm, and the contour levels
are $15~\mjybm \times -5$, $-3$, 3, 5, 7, 9, and~12.
(b)~1998 September~25.  The resolution is $30\arcsec \times 15\arcsec$.  The rms noise level is 11~\mjybm, and the contour
levels are the same as those in (a). \src\ is located in the northeast
corner of the image.}
\label{fig:discovery}
\end{figure}

Figure~\ref{fig:discovery} compares images of a region approximately
1\arcdeg\ north of Sgr~A${}^*$ from the observations acquired during 1986--1989
and in~1998 September.  Clearly evident ($\simeq 20\sigma$) in the later
image is a bright ($216 \pm 20$~mJy), unresolved source, \src.

Because \src\ appears in only a single epoch, we have conducted a
number of tests in order to establish that our detection of
\src\ is robust.  First, we verified that the source is not an imaging
or self-calibration artifact.  \src\ is visible on a dirty image,
prior to any deconvolution or self-calibration.  The source was also
present regardless of the number of facets (9 \textit{vs.} 49
\textit{vs.} 357) used in the wide-field imaging.  Its presence
regardless of the different number of facets means that it is unlikely
to be a phase error resulting from the non-coplanar nature of the
\hbox{VLA}.  Furthermore, the source's location within a facet
depended upon the number of facets used, yet the flux densities
derived from the three images are consistent.  As the magnitude of
phase errors depend typically upon the distance from the phase center
of the facet, we consider the constancy of the source's flux density
to be strong evidence that the source is real.

A second test was made by imaging sections of the data.  We split the
data in both time and frequency.  Splitting the observations into
four, approximately 2-hr time intervals, we find that the transient
flux density is constant to within their uncertainties for all but the
first time interval. However, the observing logs indicate several
problems encountered during the beginning of the run that could
account for that discrepancy.  Indeed, if we split the first interval
in half, \src\ appears in the second half but not the first half.
Splitting the observations into three equal frequency intervals, we
again find the transient flux density measurements to be
consistent. Taken together, we conclude that our detection of
\src\ is robust.

\begin{figure}
\begin{center}
\mbox{\psfig{file=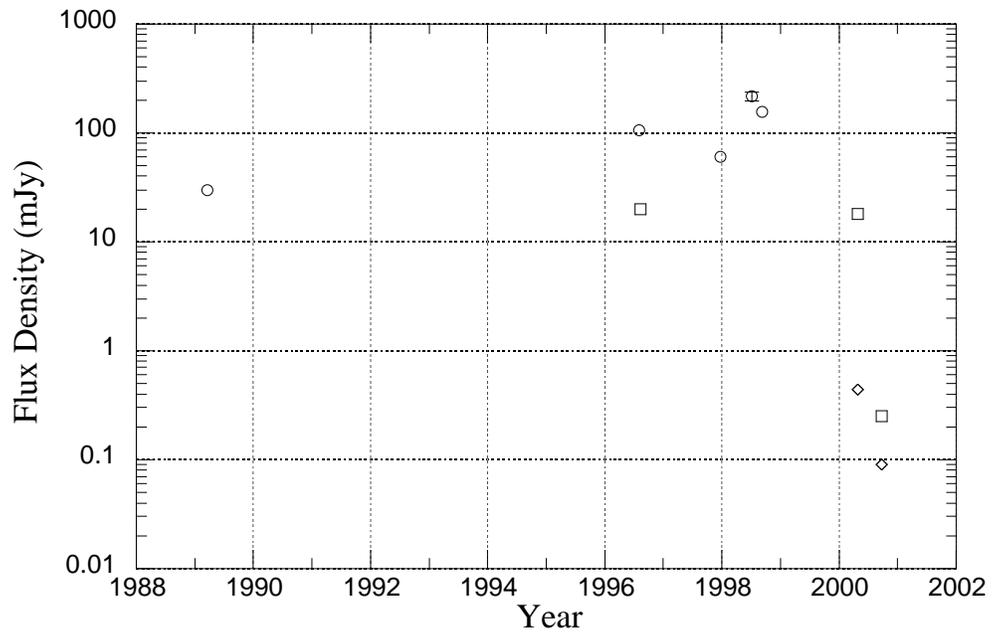,width=0.8\textwidth,silent=}}
\end{center}
\caption{Measurements of the new radio transient, \src. All data
points are $3\sigma$ upper limits except the peak at $216 \pm
20$~mJy. The circles, squares, and diamonds are measurements at~0.33,
1.4, and~4.86~GHz, respectively.}
\label{fig:ltcurve}
\end{figure}

\src\ does not appear on the image by LaRosa et al.~(2000) nor on the 1996 or
other 1998 images. A search of the All-Sky Monitor on the Rossi X-ray
Timing Explorer All-Sky Monitor (RXTE/ASM) database was kindly
performed by R.~Remillard~(2000, private communication), but no X-ray
counterpart was found, as was the case for the \hbox{GCT}.  We
obtained targeted, follow-up VLA observations at higher frequencies on
2000 July 15 and 19 and 2000 December 11. However, even with the high
resolution (0.81\arcsec\ $\times$ 0.37\arcsec) and sensitivity
(0.03~\mjybm) achieved in the latter observations at 4.86~GHz, the
transient was not recoverable. Figure~\ref{fig:ltcurve} shows the flux
density of \src\ measured on~1998 September~25 along with $3\sigma$
upper limits for each of the other epochs.

We fit an elliptical gaussian to \src.  The upper limit on its
deconvolved angular diameter corresponding to a flux density $1\sigma$
above the best-fit value is 9\arcsec.  In turn this implies a lower
limit on its brightness temperature of~$5 \times 10^{4}$~\hbox{K}.  As
this is a conservative estimate, we conclude that \src\ is nonthermal.
We have no spectral information on \src, as it was detected at only a
single frequency.  We shall assume that its spectral index was the
same as that for the GCT, $\alpha = -1.2$ ($S_\nu \propto
\nu^\alpha$).

The rapid growth and decline in brightness of \src\ is consistent with
that for the GCT and the radio counterpart to the X-ray transient
A1742$-$28.  For frequencies between~1.36~GHz and~1.67~GHz, the GCT
decreased to approximately 12\% of its peak flux density in only a
month (\cite{zhaoetal92}).  The only detection of the radio
counterpart to A1742$-$28 was made on~1975 March~30 at~0.96~GHz,
followed by non-detections on~1975 May~3 and~June~20 at~0.408
and~0.96~GHz, respectively.

Assuming that \src\ decayed according to a power law ($S_{\nu} \propto
[t-t_0]^{\beta}$) with the same index as the GCT, $\beta = -0.67$
(\cite{zhaoetal92}), and that the peak occurred no later than one day
prior to our detection, we would expect its 1.4~GHz flux density to
have been no less than about~1~mJy in~2000 December.  As the noise
level of our 2000 December~11 follow-up 1.4~GHz observations was much
less (0.08~\mjybm) than this extrapolated value, we conclude that
\src\ decayed much faster than the GCT, its spectrum was steeper, or
both.

In contrast to the GCT and A1742$-$28, both of which were within
1\arcmin\ or so of Sgr~A${}^*$, \src\ is located over 1\arcdeg\ away,
but at a low Galactic latitude ($b = 0\fdg4$).  At about 150 pc from
Sgr~A${}^*$ in projection, \src\ is almost certainly in the
Galactic bulge, if not in the GC itself.

\subsection{\xtesrc}\label{sec:xte}

A second, bright transient that we identify as the radio counterpart
to the X-ray transient \xtesrc\ (\cite{hrm98a}, 1998b; \cite{rhm98})
also appears in our images.  This radio transient reached its maximum
in~1998 June and was monitored extensively from~1998 June to October
at frequencies between~1.45 and~22.5~GHz (\cite{hjellmingetal98c};
Rupen et al.~2001, in preparation).

\begin{figure}
\begin{center}
\mbox{\psfig{bbllx=105pt,bblly=240pt,bburx=515pt,bbury=620pt,file=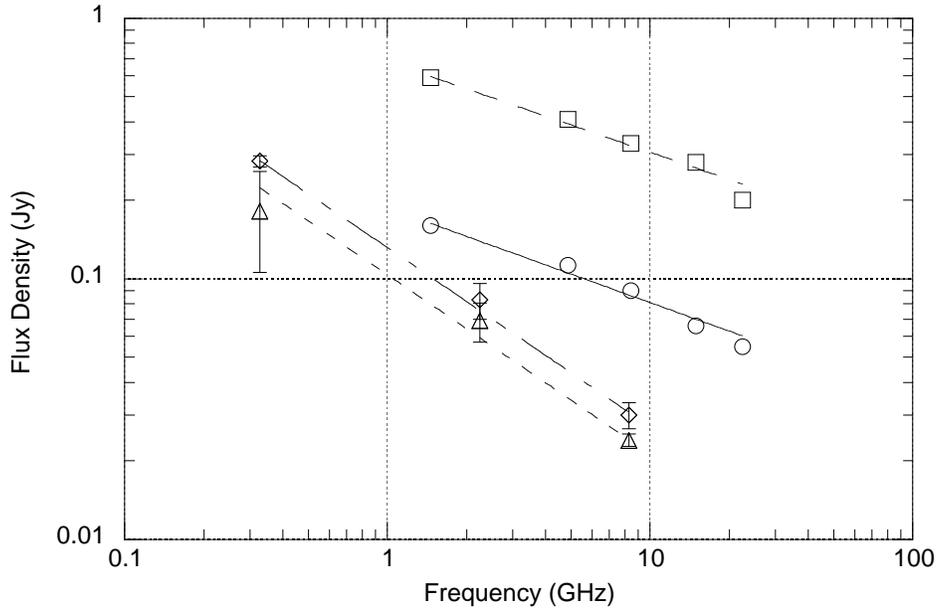,width=0.9\textwidth,silent=}}
\end{center}
\caption{Spectra of \xtesrc\ derived from measurements made on~1998
September 25 (diamonds) and November~29 (triangles).  The spectral
index on both dates is $\alpha =-0.69 \pm 0.09$.  The 2.25
and~8.30~GHz measurements for 1998 September and November were kindly
provided by M.~Rupen.  Also shown are preliminary measurements
obtained by Hjellming et al.~(1998b) on~1998 June~10.33 (circles) and
\cite{rhm98} on~1998 June~14.31 (squares). The spectra are both fit by
power laws with spectral index $\alpha \approx -0.35 \pm 0.03$.}
\label{fig:xtespectrum}
\end{figure}

We provide here the first flux density measurements of this transient
at~0.33~GHz. The flux density on~1998 September~25 was $283 \pm
14$~mJy, decreasing to $182 \pm 76$~mJy on~1998 November~29.
Figure~\ref{fig:xtespectrum} shows the spectra obtained by combining
our 0.33~GHz flux densities from~1998 September and November with flux
densities obtained by the Green Bank Interferometer (GBI) at~2.25
and~8.30~GHz on the same dates (Rupen et al.~2001, in preparation).
For both dates, the spectral indices are $\alpha = -0.69 \pm 0.09$.

Hjellming et al.~(1998a) reported preliminary flux density
measurements at~1.46 and~4.86~GHz from VLA observations on~1998
June~7.4, and Hjellming et al.~(1998b) and Rupen et al.~(1998)
reported flux density measurements at five frequencies from VLA
observations on~1998 June~10.33 and~14.31.  The spectral indices of
these higher frequency measurements are $\alpha \approx -0.4$,
considerably flatter than the value we find from our later
measurements.  We conclude that there was spectral steepening of the
source between~1998 June and September.

Assuming that the magnetic field in the emission region was of order
1~$\mu$G and using 0\farcs25 as the diameter of the source
(\cite{rhm98}), we conclude that the source was probably \emph{not}
synchrotron self-absorbed.  Thus, we estimate the 0.33~GHz flux
density on~1998 June~7, 10, and~14 to be 106, 280, and~1006~mJy,
respectively.  Additional analysis of these combined data sets will be
reported elsewhere (Rupen et al.~2001, in preparation).

\section{Conclusion and Future Work}\label{sec:conclude}

We have detected two radio transients in the Galactic center.
\begin{enumerate}
\item \src\ (Figure~\ref{fig:discovery}) was a nonthermal source with
a rapid temporal evolution.  Although limited, our spectral and
temporal constraints indicate that its characteristics were similar to 
that of previous GC radio transients, though its temporal
evolution must have been faster, its spectrum steeper, or both.  No
X-ray counterpart was detected.

\item We have also detected \xtesrc\ at a frequency for which it had not
been observed previously, and we present evidence that the spectrum
has steepened significantly over a period of a few months
(Figure~\ref{fig:xtespectrum}).
\end{enumerate}

The observations we present here, combined with previous GC
radio transients, suggest that the GC may harbor a population(s) of
radio transients.  However, our observations also indicate that the
rapid brightening and fading of transient sources requires much more
extensive and frequent observations in order to detect and monitor a
large number of radio sources. These measurements can then be used to
determine, e.g., the angular distribution and typical timescale of the
source population toward the \hbox{GC}. Such statistics and other
individual and group properties are necessary to constrain the nature
of the transient and variable source population(s).

An efficient transient monitoring program requires high-resolution,
high dynamic range, large field-of-view images.  VLA observations
at~0.33~GHz of the GC present a near optimal means to achieve these
goals.  The large field of view (2\fdg5) allows the entire GC to be
imaged with a single observation.  In its extended configuration (A
and~B), the VLA provides an angular resolution of~20\arcsec\ or better
and a sensitivity of~10~\mjybm\ or better for a 6-hr synthesis.  Also,
we are developing an algorithm to detect accurately source variations
relative to a fiducial, full-field model, with less observing time per
epoch.  Observations at~0.33~GHz also exploit the apparent
steep-spectrum nature of these sources without being affected unduly
by free-free absorption.

\acknowledgements
We thank T.~Cornwell and D.~Frail for discussions about transient
imaging, K.~Weiler for helpful discussions, and M.~Rupen for
discussions about \xtesrc.  SDH thanks S.~Bollinger
and A.~Bartleson for their assistance with the measurement and
comparison of radio sources detected at various epochs.  The National
Radio Astronomy Observatory is a facility of the National Science
Foundation operated under cooperative agreement by Associated
Universities, Inc.  The Green Bank Interferometer was a facility of
the National Science Foundation operated by the NRAO in support of
NASA High Energy Astrophysics programs.  SDH is supported by funding
from the Jeffress Memorial Trust, Research Corporation, and the Sweet
Briar College Faculty Grants program.  Basic research in radio
astronomy at the NRL is supported by the Office of Naval Research.

\end{document}